\documentclass[12pt]{l4dc2021} 
\usepackage{amsmath, amssymb, natbib, graphicx, url, algorithm2e}
\usepackage{enumitem} 
\usepackage{mathtools}

\newcommand{\hnorm}[1]{\|{#1}\|_{\mathcal H}}

\newcommand{\expected}[2]{{\mathbb E}_{#1}\ #2}
\newcommand{\rkhs}{\ensuremath{\mathcal H}\xspace}

\newcommand{\supp}[2]{\ensuremath{\delta^*_{{#1}}}({#2})} 

\newcommand{\rset}{\mathbb R}  
\newcommand{\domain}{\ensuremath{\mathcal X}\xspace}
\newcommand{\ipm}{\ensuremath{D_{\mathcal F}}\xspace} 

\newcommand{\bias}{\operatorname{Bias}} 

\newtheorem{assume}{Assumption} 

\usepackage{xcolor}

\title[Distributional Robustness Regularized Scenario Optimization]{Distributional Robustness Regularized Scenario Optimization with Application to Model Predictive Control}
\usepackage{times}

\begin{document}
\author{\Name{Yassine Nemmour} \Email{ynemmour@tuebingen.mpg.de}\\
  \Name{Bernhard Sch\"olkopf} \Email{bs@tuebingen.mpg.de}\\
  \Name{Jia-Jie Zhu} \thanks{Now at Weierstrass Institute for Applied Analysis and Stochastics (WIAS) Berlin.}\Email{jzhu@tuebingen.mpg.de}\\
  \addr Empirical Inference Department\\ Max Planck Institute for Intelligent Systems, T\"ubingen, Germany}


\maketitle

\begin{abstract}%
  We provide a functional view of distributional robustness motivated by
  robust statistics and functional analysis. This results in two practical
  computational approaches for approximate distributionally robust nonlinear optimization based on gradient norms and reproducing kernel Hilbert spaces.
  Our method can be applied to the settings of statistical
  learning with small sample size and test distribution shift.
  As a case study, we robustify scenario-based stochastic model predictive control with general nonlinear constraints.
  In particular, we demonstrate constraint satisfaction with only a small number of scenarios under distribution shift.
\end{abstract}


\section{Introduction}
This paper studies techniques for handling constraints in the form of uncertain nonlinear inequality
\begin{equation}
    \label{eq:uncertain}  
    \quad f(x, \xi) \leq 0,
\end{equation}
where $x$ is the decision variable and $\xi$ is an uncertain variable.
One principled approach to handle \eqref{eq:uncertain} is \emph{robust optimization} (RO)
\citep{soysterTechnicalNoteConvex1973,ben-talRobustOptimization2009a,bertsimasTheoryApplicationsRobust2011c}.
In RO, uncertain inequalities~\eqref{eq:uncertain} can be written as their \emph{robust counterparts} (RC)
\begin{equation}
    \label{eq:rc}  
    \text{(RC)}\quad f(x, \xi) \leq 0, \forall \xi \in \domain,
\end{equation}
where \domain is an uncertainty set where the uncertain $\xi$ lives.
The methodology then proceeds to reformulate \eqref{eq:rc} as solvable deterministic programs for useful choices of uncertainty sets \domain such as polyhedra; cf. \citep{ben-talRobustSolutionsOptimization2013}.
However, the downside of canonical RO methodology is its potential conservativeness ---  robustifying against every element in the modeled uncertainty set equally is hardly necessary in practice.

\emph{Distributionally robust optimization} (DRO) \citep{delageDistributionallyRobustOptimization2010,gohDistributionallyRobustOptimization2010,scarfMinmaxSolutionInventory1958}
extends RO techniques to probability measures while mitigating the conservativeness.
It is a promising tool to treat many robustness challenges in machine learning and data-driven control. 
DRO with various constraints have been developed, including finite-order moment bounds \citep{delageDistributionallyRobustOptimization2010,scarfMinmaxSolutionInventory1958,zymlerDistributionallyRobustJoint2013,milzApproximationSchemeDistributionally2020a},
$f$-divergence constraints \citep{ben-talRobustSolutionsOptimization2013,iyengarRobustDynamicProgramming2005,nilimRobustControlMarkov2005,wangLikelihoodRobustOptimization2016,duchiStatisticsRobustOptimization2018,bayraksanDataDrivenStochasticProgramming2015}, and, more recently, probability metrics such as the popular Wasserstein distances \citep{mohajerinesfahaniDatadrivenDistributionallyRobust2018,zhaoDatadrivenRiskaverseStochastic2018,gaoDistributionallyRobustStochastic2016, blanchetRobustWassersteinProfile2019,xieDistributionallyRobustChance2019} and kernel distances \citep{zhuKernelDistributionallyRobust2020b,xuRobustnessRegularizationSupport2009,Staib2019}. We refer to \citep{Rahimian2019} for a survey.
A significant amount of recent effort has been focused on the Wasserstein DRO due to many of its attractive properties. However, common approaches assume either a restricted function class or the knowledge of Lipschitz constant which is often inaccessible in modern applications as discussed in \citep{virmauxLipschitzRegularityDeep2018,biettiKernelPerspectiveRegularizing2019}.
In particular, this paper is interested in handling general nonlinear inequality~\eqref{eq:uncertain}, which falls outside the scope of commonly-used reformulation techniques.

The RO methodology has been widely adopted in robust constrained control and model predictive control (MPC)
\citep{bemporadRobustModelPredictive1999,Langson2004a,diehlRobustDynamicProgramming2004b,Mayne2005}
A central component of those approaches is to guarantee constraint satisfaction of the optimal control problem under uncertainty, which may be modeled by (RC)~\eqref{eq:rc}.
This is also intricately related to recent constraint tightening techniques in stochastic MPC \citep{Kohler,Bonzanini2019,Santos2019,Hewing2020a,Mark2020}; see \citep{Mesbah2016, Farina2016} for more details of the prior art.
Using tube-based stochastic MPC as an example, this paper shows how disributionally robust nonlinear optimization techniques can be applied to constraint tightening for general nonlinear constraints.
Before going further, we make an important distinction between two different settings in learning and data-driven control. 
With a possible abuse of terminology, we refer to them as \emph{statistical learning with small datasets} and \emph{distribution shift}.
\begin{enumerate}[label=(\roman*),noitemsep,topsep=0pt]
    \item In the statistical learning with small datasets setting, the training and test data are assumed to be sampled from the same generating distribution $P_\text{true}$.
    Note this is a setting commonly studied in stochastic MPC constraint tightening approaches such as~\citep{Hewing2020a,Mark2020}.
    In this setting, the empirical distribution ${\hat P}_N$ deviates significantly from $P_\text{true}$ due to the small sample size.
    Methods such as scenario approach cannot guarantee high-level of robustness.
    \item In the distribution shift setting, the test data is generated from a shifted distribution, denoted as $P_{\text{shift}}$, different from $P_\text{true}$.
    This is often the case in practice since the true system is inevitably different from the mathematical model.
    Then, even with large datasets, common stochastic optimization and control design still performs poorly during test time.
    This was the topic of \citep{Zhu2020} and is also related to the issue of adversarial robustness in machine learning and simulation-to-reality transfer in robotics.
\end{enumerate}
\paragraph{Contribution.}
\begin{enumerate}[label=(\roman*),noitemsep,topsep=0pt]
    \item
      We provide a unified functional view of distributional robustness motivated by
      robust statistics and robust optimization. Our perspective leads to two simple and practical
      computational approaches based on gradient norm and Kernel DRO. Our methods are applicable to general nonlinear constraints.
    \item
      We clearly point out two distinct settings where our distributional
      robust counterpart (DRC) framework can be useful: statistical
      learning with small sample size and test distribution shift.
      Previous works mainly focus on the former setting.
    \item
      In a case study, we robustify scenario-based stochastic tube-based MPC with general nonlinear constraint, whereas the
      previous approaches only consider linear constraint tightening. 
      Most significantly, we address a common shortcoming of the scenario approach, by achieving constraint satisfaction with only a small number of scenarios.
    \item
      While the refined rate of kernel distance in Lemma~\ref{thm:rate_mmd} has already been discovered, it could not be applied to general DRO since the function of interest does not live in a known RKHS. This paper unlocks
      this potential.
    \end{enumerate}

\paragraph{Notation.}
Depending on the context, $\| \cdot \|$ denotes the norm of a vector, matrix, or operator in the corresponding normed spaces.
$\nabla f$ denotes the gradient operator. Depending on the context, $\nabla f(x)$ denotes a gradient vector in Euclidean spaces or a G\^ateaux derivative in more general Hilbert spaces. The Minkowski set addition is given by $A \oplus B = \{a + b : a \in A, b\in B \}$. The Pontryagin set difference is defined as $A \ominus B = \{a : a + B \subseteq A\}$. We refer to the decision variable as $x \in \mathbb{R}^{n_x}$ and $\xi \in \mathbb{R}^{n_x}$ generally refers to the uncertain variable.

\section{Preliminaries}
\subsection{Reproducing kernel Hilbert space and integral probability metrics}

Recall that a kernel is similarity measure modeled by a symmetric function $k\colon \mathcal{X} \times \mathcal{X}\to \rset$.
If there exists a Hilbert space \rkhs and feature map $\phi\colon
\mathcal{X} \to \rkhs$ such that $k(x,y) = \langle \phi(x), \phi(y)
\rangle_\rkhs$ is an inner product on $\rkhs$, a space of real-valued functions on $\mathcal{X}$. 
Then \rkhs is a reproducing kernel Hilbert space (RKHS) associated with $k$.
A commonly used kernel is the Gaussian kernel $k(x,y) = \exp\left( - \frac{\|x-y\|^2_2}{2\sigma^2}
\right)$ where
$\sigma>0$ is the bandwidth parameter.

This paper concerns DRO constrained by sets described by a class of probability metrics, the \emph{integral probability metric} (IPM) \citep{mullerIntegralProbabilityMetrics1997,grettonKernelTwosampleTest2012,sriperumbudurEmpiricalEstimationIntegral2012}.
Given two probability measures $P,Q$,
an IPM generated by some function class $\mathcal F$ can be written as 
$\ipm (P,Q):=\sup_{f\in \mathcal F}\{|\int f d (P-Q)|\}$, where the optimizer $f^*$ is a witness function.
The family of IPMs include, among others, the Kantorovich metric (the dual representation of Wasserstein-1 metric), the total variation distance, and kernel distance (also known as the maximum mean discrepancy (MMD)).
In this paper, we focus on the Wasserstein-1 metric and the MMD via a functional approach, i.e., by manipulating the function $f$ in the IPM definition.
The Wasserstein-1 metric is obtained by choosing the function class
$\mathcal F=\{f:\mathrm{lip}{(f)}\leq 1\}$ where $\mathrm{lip}{(f)}$ denotes the Lipschitz constant of $f$.
On the other hand, by choosing
$\mathcal F=\{f:\hnorm{f}\leq 1\}$ where $\hnorm{\cdot}$ is the norm in an RKHS, we recover
the MMD associated with the \rkhs. Given $N$ i.i.d. samples $X_i\sim P, Y_i\sim Q$, a sample-based estimator for MMD can be computed by
$
\widehat{MMD}^2(P, Q) = \frac{1}{N(N-1)} \sum_{i\neq j}[k(X_i, X_j) + k(Y_i, Y_j) - k(X_i, Y_j) - k(X_j, Y_i)]
$.

The statistical guarantees for DRO constrained by probability metrics typically concern the size of the ambiguity set, e.g., the minimum radius $\epsilon$ of some metric-ball $B_\epsilon = \{m: D(m, \hat{P}_N)\leq \epsilon \}$ centered at the empirical distribution $\hat{P}_N$ such that it contains the unknown true generating distribution $P_{\text{true}}$.
The convergence rate for Wasserstein DRO have been established originally in \citep[Theorem~3.4]{mohajerinesfahaniDatadrivenDistributionallyRobust2018} to be $\mathcal{O}((1/n)^{\frac1d})$ where $d$ is the dimension of data. That rate suffers from the curse of dimensionality and contains unknown constants.
However, this is still an active area of research. 
We refer to recent works that report more refined rates, e.g. \citep{siQuantifyingEmpiricalWasserstein,blanchetRobustWassersteinProfile2019}.
In contrast, the refined rate for kernel distances (MMD) does not involve unknown constants or the dimensions of \domain. 
Below is due to \citep[Proposition A.1]{tolstikhinMinimaxEstimationKernel2017}.
\begin{lemma}[\cite{tolstikhinMinimaxEstimationKernel2017}]
    \label{thm:rate_mmd}
	Suppose ${\hat P}_N$ is the empirical distribution ${\hat P}_N=\sum_{i=1}^N{\frac1N}\delta_{\xi_i}$, where $\xi_i$ are i.i.d samples of the true distribution
    ${P_\text{true}}$.
    Let $C$ be a constant such that $\sup_{x\in\domain} k (x,x)\leq C<\infty$.
    Then, with probability at least $1-\alpha$,
    \begin{equation}
    \label{eq:rate_mmd}
        \operatorname{MMD}(\rkhs, \hat P_N,{P_\text{true}} )
        \leq \sqrt{\frac{C}{N}}+\sqrt{\frac{2 C \log (1 / \alpha)}{N}}.
    \end{equation}
\end{lemma}
For the commonly used  Gaussian RKHS, $C$ is easily computable by $\sup_{x\in\domain} k (x,x) = 1$,
resulting in a computable rate.
It is also possible to choose ambiguity set sizes empirically based on training data, such as bootstrap; see, e.g., \citep{grettonKernelTwosampleTest2012}.
The true power of Lemma~\ref{thm:rate_mmd} only manifests when the corresponding kernel $k$ is known, as is the case enabled by this paper's method.

\subsection{Scenario-based approaches to stochastic optimization}
In the scenario optimization literature, such as the work in \citep{Campi2011}, a chance constraint $\mathbb P (f(x, \xi) \leq 0) \geq 1 -\alpha$ is replaced by a sample-based one. Then, the uncertain inequality constraint is required to hold for every sample. \citet{Hewing2020a} combine the data-driven nature of scenario methods with the offline computation of tube-based MPC. As an application of our methodology, we will consider the MPC formulation \citep[eq. 11(a-g)]{Hewing2020a} for the remainder of the paper.
Moreover, they can make use of the bound derived in \citep[Theorem 2.1]{Campi2011}. It provides a lower bound on the number of required scenarios such that we find a solution to the chance constrained problem with at least probability $1 - \beta$. A simplified version of the scenario bound is found in \citep{Hewing2020a}
\begin{equation}
  \label{eq:stat_guarantee_scen}
  N_s \geq \frac{2}{1-\alpha}((d-1)\text{ln}(2) - \text{ln}(\beta)),
\end{equation}
where $N_s$ is the number of scenarios and $d$ is the dimensionality of the decision variable. For high satisfaction probabilities and complex problems, i.e., large decision variables, that still suffers from the curse of dimensionality.
Recent works investigate regularization in scenario-based approaches, addressing conservativeness and better generalization. As previously stated, penalizing every scenario equally, does increase conservativeness. \citet{Campi2013} use $\ell 1$ regularization, in form of an $\ell 1$ constraint for the min-max formulation of the scenario-based approach, to encourage sparsity of the decision variable. In \citep{formentin_tuning_2017} constraint relaxation is approached by reducing the number of active scenario constraints. They achieve this with $\ell 2$ regularization. The work in \citep{Zhu2020a} reduces conservativeness by removing scenarios as long as the kernel mean embedding (KME) of the reduced scenario set does not differ significantly from the KME of all scenarios. That can be formulated as a convex optimization problem and solved efficiently.
\section{Theory and Methods}
\subsection{Distributionally robust counterparts}

\begin{figure}[tb!]
  \floatconts{}{\caption{The dashed line in (a) is the support of the probability distribution. The ambiguity set of the scenario approach is colored in pink.
  Intuitively, our way to robustify is a polytope with round corners in the space of probability measures.
  We can choose the size of $B$, in blue, according to statistical learning bounds such as in Lemma~\ref{thm:rate_mmd}. This way, the size of B is large when $N$ is small and shrinks as $N$ increases, while the ambiguity set, yellow area, converges to the dashed line. Figure (b), from \citep{zhuKernelDistributionallyRobust2020b}, visualizes the intuition of the majorization in Proposition \ref{thm:kdro}.}}
  {\subfigure[][b]{
    \label{fig:sl_robustify}
    \centering
    \includegraphics[height=0.2\textwidth]{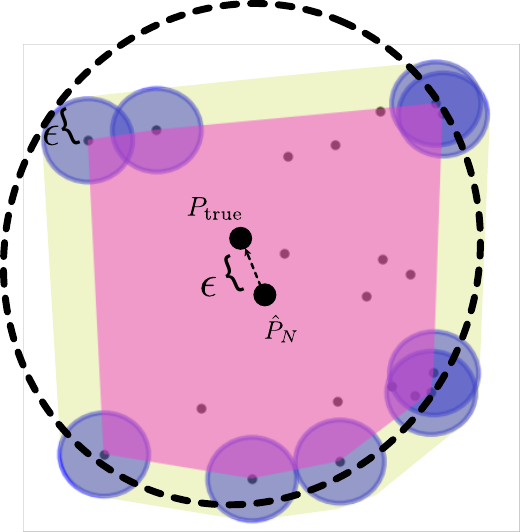}
  }
  \subfigure[][b]{
    \label{fig:kdro_intuition}
    \includegraphics[width=0.45\textwidth]{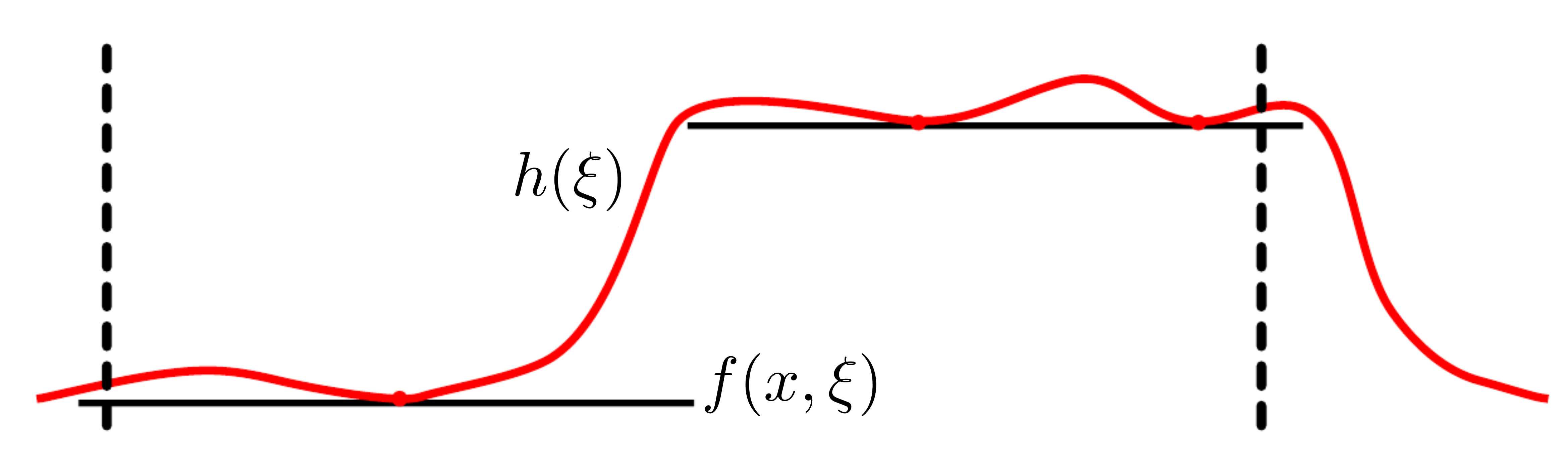}
  }}
\end{figure}

The starting point of our theory is the distributionally robust counterpart (DRC) of the uncertain nonlinear inequality in \eqref{eq:uncertain} and \eqref{eq:rc}
\begin{equation}
    \label{eq:drc}
    \text{(DRC)}\quad \sup_{P\in K}\mathbb E _ P f(x, \xi) \leq 0, \forall P \in K,
\end{equation}
where $K$ is a set of probability distributions, known as the ambiguity set.
(DRC) describes the worst-case expected value of the constraint function depending on an uncertain variable $\xi$.
Throughout the paper, we assume the following non-restrictive conditions for the (DRC).
\begin{assume}
    \label{thm:assume}
    $f(x, \cdot)$ is upper semicontinuous and $f(x, \xi)>-\infty$ for at least some $\xi\in\domain$.
    $K$ is closed convex and non-empty.
\end{assume}
Many existing formulations can be written as special cases of (DRC), such as the canonical worst-case RO, sample average approximation. 
\citet{zhuKernelDistributionallyRobust2020b} give an exhaustive list of the special cases. One special case of (DRC) is the polytopic constraints, e.g., by letting $f(x, \xi) = H (x+\xi) - b$, which is considered in \citep{Hewing2020a,Mark2020}.
Those works then used polytopic probabilistic reachable sets (PRS) and Pontryagin difference for constraint tightening. 
In this paper, we are interested in the more general classes of (DRC).
We also note that such functions $f$ may fall outside the scopes of typical Wasserstein DRO reformulation techniques such as \citep{MohajerinEsfahani2018}. Hence, this paper's investigation also constitutes a generalization in that regard.
We now turn to a probabilistic robustness method --- the scenario approach to stochastic optimization.
\begin{example}[Scenario approach as DRC]
    Let $K = \mathrm{conv}(\delta_{\xi_1}, ..., \delta_{\xi_N})$.
Then (DRC) is given by
$$
\sup_{P\in K} \mathbb E_P f(x, \xi) = \max_i f(x, \xi_i)\leq 0,
$$
which is the scenario approach. It can be seen as a special case of (DRC) where the ambiguity set $K$ is chosen to be a data-driven polytope $\mathrm{conv}(\delta_{\xi_1}, ..., \delta_{\xi_N})$.
\end{example}
Let us consider the setting where $N$ is small and the bound in \citep{Campi2011} does not provide robustness anymore.
We propose to use the following \emph{Minkowski sum}
to robustify the scenario approach in this setting.
We illustrate our idea in figure~\ref{fig:sl_robustify}.
\begin{example}[Constraint tightening via Minkowski sums of ambiguity sets]
    \label{ex:minkowski}
    Let $K$ be the ambiguity set specified by the polytope $K=\mathrm{conv}(\delta_{\xi_1}, ..., \delta_{\xi_N})$ in the scenario approach. 
    Let $B$ be some convex set in the topological vector space of signed measures.
    Then, we have the following
    \begin{multline}
        \label{eq:minkowski}
    \sup_{P\in (K+B)\cap \mathcal P} \mathbb E_P f(x, \xi) \leq \sup_{P\in K} \mathbb E_P f(x, \xi)+\sup_{P\in B} \mathbb E_P f(x, \xi)=  \max_i f(x, \xi_i) + \supp{B}{f(x, \cdot)} \leq 0,
    \end{multline}
    where $\supp{B}{f} $ denotes the support function given by $\supp{B}{f} \coloneqq \sup_{\mu \in B} \mathbb E_\mu f(x, \xi)$.
\end{example}
Compared with the scenario approach, there is an extra tightening term $\supp{B}{f}$. In the next section, we provide approaches to compute the tightening by reformulating (DRC).
\subsection{A functional view of distributional robustness}
\label{sec:functional}
We now take a functional approach commonly adopted in robust statistics \citep{fernholzMisesCalculusStatistical2012,hampelRobustStatisticsApproach2011,misesAsymptoticDistributionDifferentiable1947} and also in stochastic programming, see, e.g., \citep[Chapter~6.6.3]{shapiroLecturesStochasticProgramming2014}.
This view offers a clear picture of the role that function spaces play in distributional robustness.
We view the risk measure as a function of the probability measure.
Let $P, \hat P$ be probability measures.
With a slight abuse of terminology, we define the \emph{bias} as
\begin{equation}
    \label{def:bias}  
    \bias (P; f, \hat P) := \int{f}\ d{(P-\hat P)}.
\end{equation}
Bias describes the risk deviation under a new distribution $P$ from the nominal one $\hat P$.
Using this quantity, we can write the (DRC) as
\begin{equation}
    \label{eq:bias_decomp}
    \sup_{P\in K}\mathbb E _ P f(x,\xi) = \mathbb E _ {\hat P} f(x,\xi) + \sup_{P\in K} \bias{(P; f, \hat P)} \leq 0, \forall P \in K.
\end{equation}
In robust statistics, $\sup_{P\in K} \bias{(P; f, \hat P)}$ is also referred to as the \emph{maxbias} or \emph{supremum bias}. It is also referred to as the regularizer.
Given fixed $\hat{P}_N$,
the decomposition in \eqref{eq:bias_decomp} tells us that the key quantity to controlling the distributional robustness is the supremum bias.
Let us see this through concrete examples from the DRO literature.
Using the functional view in \eqref{def:bias}, the examples below are straightforward consequences of
the definitions for the corresponding probability metrics.
\begin{example}[Wasserstein-$1$ metric]
    Suppose the function $f$ is $L_f$-Lipschitz continuous on the domain \domain. Let $ \operatorname{W_1}$ be the {Wasserstein-$1$} metric. Then,
\begin{equation}
    \label{eq:wsd}
    \sup_{\operatorname{W_1}(P, \hat P)\leq \epsilon } \bias{(P; f, \hat P)} = \epsilon\cdot L_f.
\end{equation}
\end{example}
As the Wasserstein-$1$ metric belongs to a larger family of the \emph{integral probability metrics} (IPM), we give another instance of this family.
\begin{example}[Maximum mean discrepancy (MMD) associated with an RKHS]
    Suppose the function $f\in\mathcal H_f$ belongs to the RKHS $\mathcal H_f$. 
    Let $ \operatorname{MMD}$ be the MMD associated with $\mathcal H_f$.
    Then,
\begin{equation}
    \label{eq:mmd}
    \sup_{\operatorname{MMD}(\rkhs_f, P, \hat P)\leq \epsilon } \bias{(P; f, \hat P)} = \epsilon\cdot  \|f\|_{\mathcal H_f}.
\end{equation}
\end{example}
Unfortunately, a limitation of the above approaches is that the Lipschitz constant $L_f$ or the associated RKHS norm $\|\cdot \|_{\rkhs_f}$ is often not accessible in general nonlinear programming and learning tasks.
It is henceforth valuable to derive a generally applicable approach.
\subsection{A simplified approach via gradient norm regularization}
\label{sec:grad}
Given a real-valued continuously differentiable function $f$ with Lipschitz constant $L_f$, we have the straightforward relationship
$
L_f \geq \sup_{x\in\domain}\|\nabla f(x)\|,
$
where equality is attained with further conditions on \domain.
This motivates the following practical quantity to control the distributional robustness without the access to a Lipschitz constant via the gradient-based lower bound to \eqref{eq:wsd}.
$$
\sup_{\operatorname{W_1}(P, \hat P)\leq \epsilon } \bias{(P; f, \hat P)} = \epsilon \cdot L_f 
\approx
\epsilon \cdot \max_i\|\nabla f(x_i)\|.
$$
We note that this optimistic lower bound is exact for constraint tightening in MPC with polyhedral constraints, such as in \citep{Hewing2020a}.
\begin{example}
Suppose $f$ is an affine function denoted by $f(x)=h\cdot x - b$, where $h\in \rset^{n_x}, b\in\rset$.
Then
$$
L_f=\sup_{x\in \mathcal X}\|\nabla f(x)\| = \|h\|. 
$$
\end{example}
We now propose a regularized scenario approach using the discussion above.
\begin{example}[Gradient norm regularized distributionally robust scenario approach]
    In the same setting as Example~\ref{ex:minkowski},
    we write the following
    gradient-norm approximate distributionally robust counterpart reformulation for constraint tightening.
    \begin{equation}
        \label{eq:grad_adrc}  
         \max_{i}f(x,\xi_i) + \epsilon\max_i\|\nabla f(\xi_i)\|\leq 0
    \end{equation}
\end{example}
We are certainly not the first to use gradient norm regularization. 
There is a large body of literature in machine learning that uses similar methods to improve robustness; see \citep{biettiKernelPerspectiveRegularizing2019} and references therein.
However, our perspective here is that the use of gradient norm regularization dates back to even earlier works in robust optimization and optimal control, for example, in \citep{diehlApproximationTechniqueRobust2006,nagyOpenloopClosedloopRobust2004a}.
Compared to those (RC) approximation techniques
this paper is interested in the distributional robustness measure instead of worst-case robustness.

\begin{remark}[Tightening for constraints learned from data]
    \label{rem:SVM}
    It is sometimes not practical to explicitly encode all the real-world constraints in control problems.
    Instead, they can be learned from data. 
    Such learned constraints are typically nonlinear which cannot be used with many existing DRO reformulation methods.
    For example, support vector machines (SVM) have been applied in the context of data-driven stochastic MPC, e.g., in \citep{shangChanceConstrainedModel2018}.
    Suppose $C$ denotes the (nonlinear) smooth decision function of a learned constraint, i.e., the constraint set is given by
    $\{\xi: C(x, \xi)\leq 0 \}$, where $x$ is a decision variable and $\xi$ corresponds to the data samples.
    We can use the proposed gradient-based tightening $\max_i C(x,\xi_i) + \epsilon\max_i\|\nabla C(\xi_i)\|\leq 0,$
    where $\epsilon$ can be chosen as the Wasserstein radius in \eqref{eq:wsd}.
    This approach is applicable to many general learning models, such as neural networks.
    
    Moreover, if $C(x, \cdot)$ is an RKHS function associated with a known kernel, as is the case of SVM, \eqref{eq:mmd} implies the (DRC) reformulation
    $
    \max_i C(x,\xi_i) + \epsilon \|C\|_{\rkhs}\leq 0
    $,
    where $\epsilon$ can be chosen as the MMD radius in \eqref{eq:mmd}.
    This tightening has an elegant interpretation of RKHS feature space perturbation discussed in \citep{xuRobustnessRegularizationSupport2009},
    but only applies to kernelized models such as SVMs.
\end{remark}

\subsection{Kernel DRC of uncertain (nonlinear) inequalities}
This section provides an approach that
enjoys wide-applicability to general constraints and the elegance of RKHS norm based tightening using the \emph{Kernel DRO} framework of \citep{zhuKernelDistributionallyRobust2020b}. It introduces a dual form of general DRO problems. The dual problem optimizes a RKHS function $h$, preserving convexity of the objective function $f$, while guaranteeing that $h$ majorizes $f$ wherever required, see \ref{fig:kdro_intuition} for an intuition.
Using the functional view in Section~\ref{sec:functional}, their result implies
    $$
    \sup_{\operatorname{MMD}(\rkhs, P, \hat P)\leq \epsilon } \bias{(P; f, \hat P)} =  \inf_{h\geq f}  \|h\|_{\rkhs}\cdot \operatorname{MMD}(\mathcal \rkhs, P, \hat P).
    $$
Compared to \eqref{eq:mmd}, we can simply choose $\mathcal H$ to be commonly used RKHSs, e.g., Gaussian RKHS, and $f$ needs not be an RKHS function. We make our statement precise in the following proposition, which adapts \citep[Theorem~3.1]{zhuKernelDistributionallyRobust2020b} to (DRC).
\begin{proposition}[Kernel DRC]
    \label{thm:kdro}
    Suppose Assumption~\ref{thm:assume} is satisfied.
    Let $\mathcal H$ be any RKHS and the set $\mathcal C \subseteq \rkhs$ be defined by $\mathcal C = \{\mu : \int \phi dP = \mu, P \in K\}$.
    Then, $x$ satisfies (DRC)~\eqref{eq:drc} if there exists $h\in\rkhs$ such that
    \begin{equation}
    \label{eq:kdrc}  
    \text{(Kernel DRC)}\quad \supp{\mathcal C}{h} \leq 0,\ f (x,\xi)\leq h(\xi) \quad \forall \xi \in \domain.\\
    \end{equation}
    where $\supp{\mathcal C}{h} \coloneqq \sup_{\mu \in \mathcal C} \langle h, \mu \rangle_{\rkhs}$ is the support function.
\end{proposition}
The proof is shown in the appendix.
The reformulation in Proposition~\ref{thm:kdro} is convexity preserving: if the inequality in (DRC) is convex in the decision variable $x$, so is \eqref{eq:kdrc}.
Proposition~\ref{thm:kdro} provides a flexible tool to handle (DRC) for general nonlinear functions $f$.
For different choices of $K$, $\supp{\mathcal C}{h}$ can be computed in closed form; see \citep[Table~1,3]{zhuKernelDistributionallyRobust2020b} and \citep{Ben-Tal2014a} for a list of expressions. 
As a concrete example, we now use Proposition~\ref{thm:kdro} in the setting of scenario approach constraint tightening to account for the distribution shift.
\begin{corollary}[RKHS norm regularized distributionally robust scenario approach]
  \label{cor:kdr_mpc}
  Suppose the ambiguity set in (DRC)~\eqref{eq:drc} is the Minkowski sum $K=\mathrm{conv}(\delta_{\xi_1}, ..., \delta_{\xi_N}) \oplus B_\epsilon$, where $B_\epsilon= \{P: \sup_{\hnorm{f}\leq 1} \expected{P}{f} \leq \epsilon \}$.
  Then, under the same assumption as Proposition~\ref{thm:kdro},  $x$ satisfies (DRC)~\eqref{eq:drc} \emph{if} there exists $h\in\rkhs$ such that
  $$
  \begin{aligned}
      \max_i h(\xi_i) + \epsilon\|h\|_{\mathcal{H}}\leq 0,\quad f (x, \xi)\leq h(\xi),\forall \xi \in \domain.
  \end{aligned}
  $$
\end{corollary}
\section{Experiments}
\label{sec:exp}
In this section we report numerical studies of earlier derived robust reformulations on a double integrator system. We compare our derived MPC methods, gradient regularization and kernel distributionally robust MPC (KDR-MPC), in settings of distribution shift and learning with small datasets. The uncertain inequality constraint is referred to as $C(x, \xi) \leq 0$. We refer to \citep[(11)(a-g)]{Hewing2020a} for the detailed MPC formulation. Note that $x$ here corresponds to the nominal state $z$ and $\xi$ to the error state $e$ in that MPC formulation. For details about the implementation we refer to the appendix in extended online version of this paper.
\paragraph{Double Integrator}
We consider the double integrator with quadratic cost that regulates the system to the origin. The dynamics are given by
\begin{equation}
\label{eq:double_pend}
x(k+1) =
\begin{bmatrix}
1 & 1\\
0 & 1  
\end{bmatrix} x(k) +
\begin{bmatrix}
  0.5 \\
  1
\end{bmatrix} u + w,  
\end{equation}
where $x(k) \in \mathbb R ^ 2$ is the state, $u \in \mathbb R$ is the control variable and $w \sim \mathcal N(0, 0.1 \cdot I)$ is the additive noise.
We consider the simple regulator problem with multiple different constraints and cost $l(x, u) = x^TQx + u^TRu$, with $Q=I$ and $R=1$.
Not only do we treat linear constraints but also other nonlinear constraints, such as an exponential constraint and a learned constraint represented by a SVM, see Remark \ref{rem:SVM}, i.e.,
\begin{equation}
\label{eq:constraints}
  C_{lin}(x, \xi) = |x_2 + \xi_2| - 3 \leq 0, \hspace{0.5cm}  C_{exp}(x, \xi) = -5 + e^{0.1(x_1 + \xi_1)} - x_2 - \xi_2 \leq 0.
\end{equation}
\paragraph{Experimental setup}
We present results on the double integrator in the case of model-mismatch. By adding a constant offset to the system matrix $\tilde{A} = A + \lambda  \mathbb I$, we can vary model-mismatch. Further, we use \eqref{eq:stat_guarantee_scen} to compute the desired number of scenarios in our setting.
For the empirical results reported in Table \ref{tab:emp_eval}, the chance constraint satisfaction level is chosen to be $\alpha=0.85$ and $\beta = 0.15$, resulting in 34 scenarios.

\begin{figure}[t]
	\floatconts
		{}
		{\caption{Comparison of scenario MPC and KDR-MPC on the double integrator with different constraint functions \eqref{eq:constraints}, linear constraint in (a), SVM constraint in (b) and exponential constraint in (c) and (d). The orange lines show the scenario approach and black refer to our KDR-MPC. The dashed lines visualize the nominal solutions against the noisy realization shown as a solid line. In (a) and (c) the brown shaded are corresponds to the tightening from the scenario approach and the blue area shows the additional tightening due to our regularization.
		The additive noise $w \sim \mathcal N(0, 0.2 \cdot I)$ has slightly increased variance in (c).
	  In (d) we verify the regularization with $\epsilon_{verify} = 0.55$, according to \eqref{eq:eps_verify}.}}
		{\subfigure[Linear constraint][b]{
			\label{fig:tube_tightening}
			\includegraphics[width=0.45\textwidth]{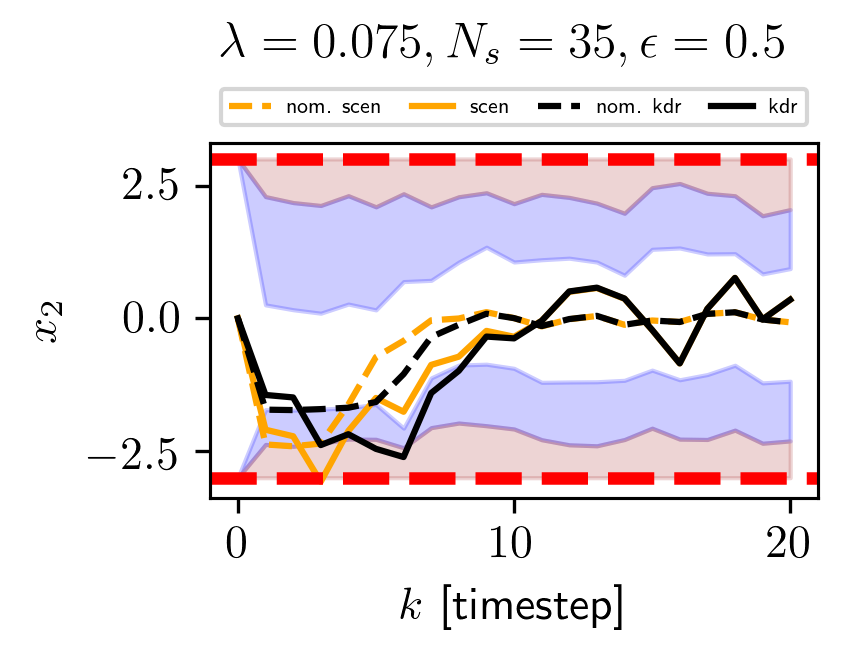}
		}
		\subfigure[SVM constraint][b]{
			\label{fig:svm_constraint}
			\includegraphics[width=0.45\textwidth]{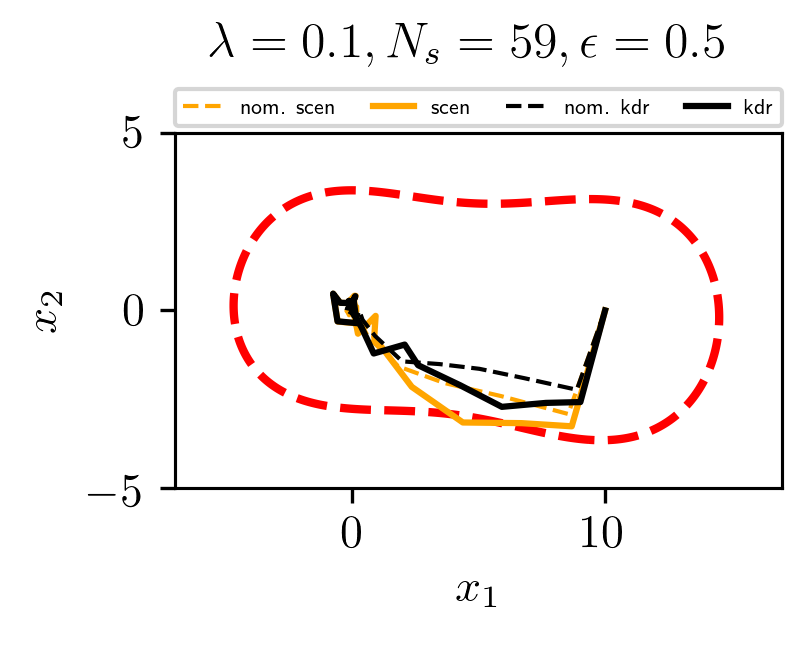}
		}
		\subfigure[No distribution shift, small sample size]{
			\label{fig:small_sample}
			\includegraphics[width=0.45\textwidth]{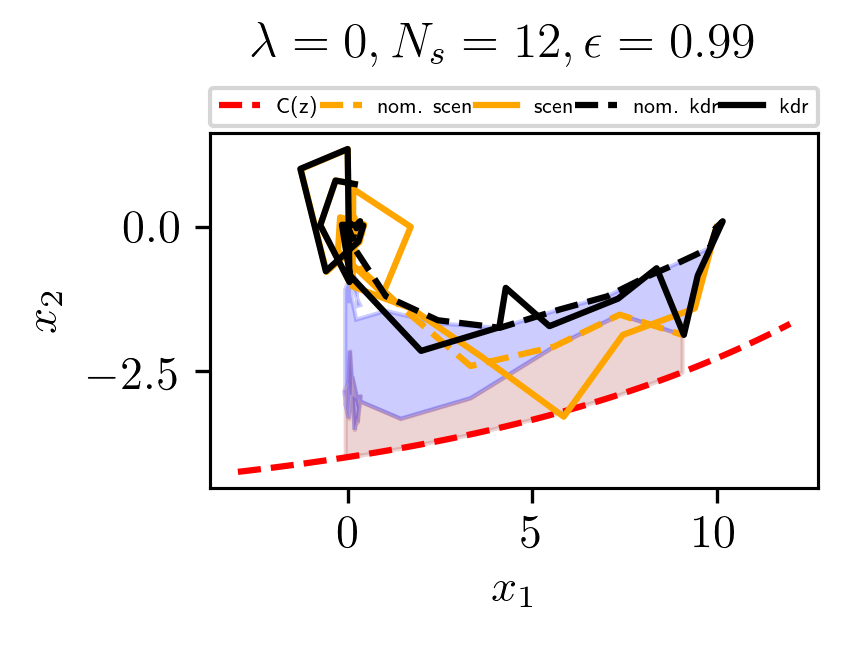}
		}
		\subfigure[Kernel DRC]{
			\label{fig:kdr_mpc}
			\includegraphics[width=0.45\textwidth]{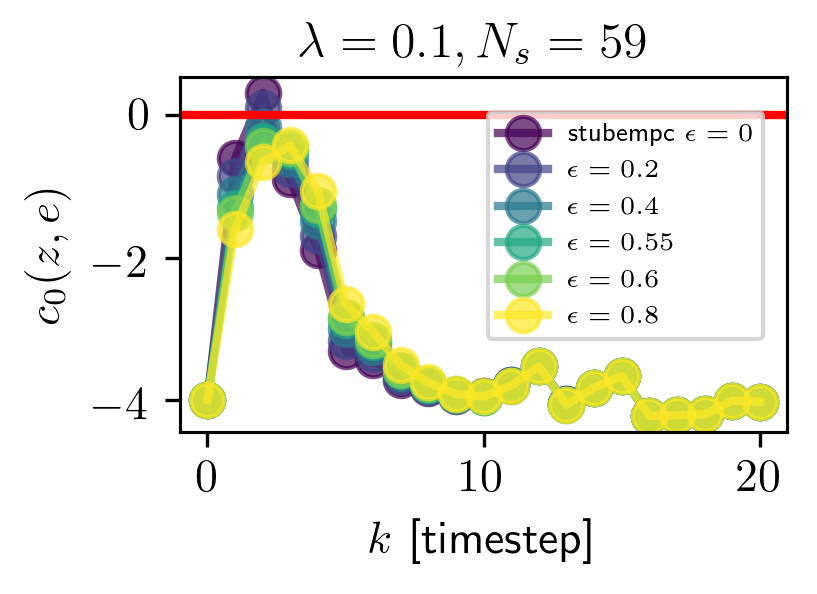}
		}
		}
\end{figure}

\begin{table}
	\floatconts{tab:emp_eval}
		{\caption{Closed-loop MPC constraint satsifaction evaluation}}
		{\begin{tabular}{c c c}
			\hline
			Method & Empirical Probability $\lambda = 0.075$ & Empirical Probability $\lambda = 0.1$\\
			\hline
			Stochastic Tube MPC & 0.907 & 0.812\\
			Gradient regularization & 0.999 & 0.994\\
			Kernel DRC & 0.999 & 0.996\\
			\hline
		\end{tabular}}
	\end{table}
\paragraph{Small sample sizes}
In Figure \ref{fig:small_sample}, we also show results when using Lemma \ref{thm:rate_mmd} to robustify in small sample-size settings. See appendix for more details.

\paragraph{Gradient regularization}
As mentioned earlier in Section \ref{sec:grad}, gradient regularization is motivated by the lower bound on the Lipschitz constant and can be used to robustify our MPC problem against shifts in distributions.
Setting $\epsilon$ to 0 in \eqref{eq:grad_adrc}, we recover the regular scenario approach.

\paragraph{KDR-MPC}
Following Proposition \ref{thm:kdro}, we can reformulate constraints as the distributionally robust counterpart. We enforce the function $f(\cdot)$ to majorize $C(x, \cdot)$ everywhere.
The RKHS function $f(\xi)$ takes the from of $f(\xi) = f_0 + \sum_i \alpha_i k(sv_i, \xi)$, where  $\alpha_i$ and $f_0$ are decision variables and $sv_i$ are the scenarios.
Figure \ref{fig:kdr_mpc} shows KDR-MPC for various robustness levels with the exponential function constraint from above. We can estimate the radius $\epsilon$ of the $\operatorname{MMD}$-ball, using the bound in \eqref{thm:rate_mmd} and an estimate of the $\operatorname{MMD}$ between the train set $X_\text{train} \sim P_\text{true}$ and the test set $X_\text{test} \sim P_\text{shift}$. Then, we get 
\begin{equation}
\label{eq:eps_verify}
\epsilon_{verify} = \sqrt{\frac{1}{N}}+\sqrt{\frac{2 \log (1 / \alpha)}{N}} + \widehat{\operatorname{MMD}}(P_{\text{true}}, P_\text{shift}).
\end{equation}

\paragraph{Learned SVM constraint}
One can frame the left-hand-side of the constraints as the \emph{decision function} of a classification problem, i.e.,
\begin{equation}
  y = +1, \text{ if } C(x) \leq 0, \quad y = -1, \text{ otherwise}.  
\end{equation}
For example, in SVM classification, the kernelized decision function is given by $C(x)=\alpha^T\phi(x) + b$, where $\alpha$ is a vector, $\phi(x)$ are the features associated with the kernel and $b$ is a constant.
We follow the (DRC) reformulation of the SVM constraint as detailed in Remark \ref{rem:SVM}. The tightening term involving the RKHS norm $\|C\|_{\mathcal H}$ can be computed offline. In that experiment, we create an artifical dataset to learn the SVM. An illustration of an SVM constraint is found in figure \ref{fig:svm_constraint}.
\section{Discussion}

Future directions of this paper include the application of our methods to nonlinear stochastic MPC as well as learning-based MPC.
Notably, coupling our (Kernel DRC) approach with the convergence result in Lemma~\ref{thm:rate_mmd} is a promising direction for new statistical guarantees.
Another direction is to design tailored numerical methods for DRO in MPC, instead of the off-the-shelf solvers used in our current experiments.

\acks{We thank Heiner Kremer for helpful feedback and discussions.}
\newpage
\bibliography{ref}
\newpage
\appendix
\section{Proofs}

Recall that $K$ is a set of probability distributions and $\mu_P$ is the kernel mean embedding of the distribution $P$ given by $\mu_P \coloneqq \int k(x, \cdot) dP$, where $k(\cdot, \cdot)$ is the kernel associated with $\rkhs$ an arbitrary RKHS. Let $\mathcal C \subseteq \rkhs$ be $\mathcal C = \{\mu_P: \int k(x, \cdot) dP, P \in K\}$. Because of assumption \eqref{thm:assume} its support function is $\delta^*_K(h) \coloneqq \sup_{\mu_P \in \mathcal{C}} \langle \mu_P, h\rangle_{\rkhs}$ for $h \in \rkhs$. The reproducing property then allows us to write the expectation of a function $h$ as $\mathbb E_{\xi \sim P}[h(\xi)] = \langle \mu_P, h\rangle_{\rkhs}$.

\begin{proof}{: Proposition \ref{thm:kdro}}\\
Suppose there exists $h \in \rkhs$ such that
\begin{alignat}{2}
& &\delta^*_K(h) &= \sup_{\mu_P\in \mathcal C} \langle \mu_P, h \rangle \leq 0 \\
& &\quad f(x, \xi) &\leq h(\xi), \quad \forall \xi \in \domain.
\end{alignat}
Note that this is the dual problem of \eqref{eq:drc} according to the Generalized Duality Theorem in \cite{zhuKernelDistributionallyRobust2020b}, where we have that the dual problem admits the same solution as the primal problem due to strong duality. Then, by the reproducing property we have
\begin{equation}
  0 \geq \sup_{P \in K} \mathbb E_P h(\xi) \geq \sup_{P \in K} \mathbb E_P f(x, \xi), \quad \forall \xi \in \domain.
\end{equation}
\end{proof}
\begin{remark}   
  The theory behind Proposition~\ref{thm:kdro} is the separation theorems of convex sets in topological vector spaces.
  Other consequences of the separation theorems include the celebrated S-Lemma (S-Procedure) in robust control and Farkas Lemma in optimization.
  We refer to \citep{polikSurveySLemma2007} for readers who are not familiar with those theorems.
  Proposition~\ref{thm:kdro} can be viewed as a generalization of those theorems in that it searches for an RKHS function $h$ rather than a real vector or scalar.
  Our idea is also reminiscent to the \emph{integral quadratic constraints}~\citep{megretskiSystemAnalysisIntegral1997} in robust nonlinear control in that we replace nonlinear functions $f$ with RKHS functions $h$ whose growth can be bounded by the computable norms.
\end{remark}

\begin{proof}{: Corollary \ref{cor:kdr_mpc}}\\
Recall that $K=\mathrm{conv}(\delta_{\xi_1}, ..., \delta_{\xi_N}) \oplus B^1_\epsilon=B^2_\epsilon \oplus (\mathrm{conv}(\delta_{\xi_1}, ..., \delta_{\xi_N}) - \hat{P})$, where $B^1_\epsilon= \{P: \sup_{\hnorm{f}\leq 1} \expected{P}{f} \leq \epsilon \}$, $B^2_\epsilon= \{P: \operatorname{MMD}(\mathcal \rkhs, P, \hat P) \leq \epsilon \}$ and $B^1_{\epsilon} = B^2_{\epsilon} - \hat{P}$. This equation holds because we only shift the sets by $\hat P$ accordingly. Let's start with equation \eqref{eq:drc}.

\begin{equation}
  \label{eq:cor4proof}
  \sup_{P\in K} \mathbb E_P f(x, \xi)\leq \sup_{P\in K} \mathbb E_P h(\xi) \leq \sup_{P \in B_{\epsilon}} \mathbb E_P h(\xi) + \sup_{P\in \mathrm{conv}(\delta_{\xi_1}, ..., \delta_{\xi_N}) - \hat P} \mathbb E_P h(\xi)
\end{equation}
We can then decompose the first term on the right-hand side above as
$$
\sup_{P \in B_{\epsilon}} \mathbb E_P h(\xi) = \epsilon \langle \frac{h}{\hnorm{h}}, h\rangle_\rkhs + \mathbb E_{\hat P} h(\xi),
$$
using the reproducing property of the RKHS and support calculus. The right term on the right-hand side can simply be rewritten as
$$
\sup_{P\in \mathrm{conv}(\delta_{\xi_1}, ..., \delta_{\xi_N}) - \hat P} \mathbb E_P h(\xi) = \sup_{P \in \mathrm{conv}(\delta_{\xi_1}, ..., \delta_{\xi_N})} \mathbb E_P h(\xi) - \mathbb E_{\hat P} h(\xi) = \max_i h(\xi_i) - \mathbb E_{\hat P} h(\xi).
$$
If we then plug in the two reformulations on the right-hand side in \eqref{eq:cor4proof} we obtain
$$
\sup_{P\in K} \mathbb E_P f(x, \xi)\leq \sup_{P\in K} \mathbb E_P h(\xi) \leq \epsilon \langle \frac{h}{\hnorm{h}}, h\rangle_\rkhs + \mathbb E_{\hat P} h(\xi) + \max_i h(\xi_i) - \mathbb E_{\hat P} h(\xi) = \epsilon \hnorm{h} + \max_i h(\xi_i)
$$
Therefore, we have that \eqref{eq:drc} holds if there exists $h \in \rkhs$ such that \eqref{cor:kdr_mpc} is satisfied.
\end{proof}

\section{Miscellaneous}
\subsection{Alternative view of Corollary \ref{cor:kdr_mpc}}
\begin{corollary}[RKHS norm regularized distributionally robust scenario approach]
  \label{cor:kdr_mpc_inverse}
  Suppose the ambiguity set in (DRC)~\eqref{eq:drc} is the Minkowski sum $K=B_\epsilon \oplus (\mathrm{conv}(\delta_{\xi_1}, ..., \delta_{\xi_N}) - \hat{P})$, where $B_\epsilon= \{P: \operatorname{MMD}(\mathcal \rkhs, P, \hat P) \leq \epsilon \}$, $P$ is a probability measure and $\hat{P}$ is the empirical distribution.
  Then, under the same assumption as Proposition~\ref{thm:kdro},  $x$ satisfies (DRC)~\eqref{eq:drc} \emph{if} there exists $h\in\rkhs$ such that
  $$
  \begin{aligned}
      \max_i h(\xi_i) + \epsilon\|h\|_{\mathcal{H}}\leq 0,\quad f (x,\xi)\leq h(\xi),\forall \xi \in \domain.
  \end{aligned}
  $$
\end{corollary}

\section{MPC Implementation}
Tube-based MPC, introduced in \citep{Langson2004a}, is one widely used application of RO in control. Stochastic MPC follows a less conservative approach than its robust counterpart. Tube-based methods separate the dynamics $x(k)$ into a nominal, disturbance-free part $z(k)$ and error-dynamics $e(k)$\footnote{$x(k)$ differs from the earlier used decision variable $x$. $z(k)$ corresponds to the $x$ commonly used in DRO literature and $\xi$ to $e(k)$.}. When considering LTI-system and for the following control parametrization $u(k) = Ke(k) + v(k)$, the error dynamics $e(k) = x(k) - z(k)$ evolve according to an autonomous system. This separation facilitates the computation of invariant sets of the error dynamics, as introduced in \citep{Blanchini1999} or more recent variations of it. The invariant sets are then used to constrain the nominal state $z(k)$, such that $z(k) + e(k)$ satisfy the original constraint on $x(k)$ hence the name tube-based. Since $e(k)$ is now an autonomous system, the tightening can be computed offline.
Contrary to common MPC methods, they consider indirect feedback, see \citep{Hewing2018}. Instead of setting the initial state to the most recent observation, the initial condition differs as they have feedback from the state measurement only through the objective function. They initialize the nominal state to the first prediction from the last MPC iteration,
\begin{equation}
  z_0(k) = z_1(k-1),
\end{equation}
where $k$ is the system time index and the subscript denotes the index inside the MPC-loop. This formulation facilitates the tube-based formulation and allows offline computation of the tube. 

We use CasADi \citep{Andersson2019} in our implementation of the optimization problem and Ipopt \citep{wachter_implementation_2006} for all numerical optimization.

\subsection{Grid point choice}

In order to enforce robustness, we impose the following inequality from corollary \ref{cor:kdr_mpc} $f(x, \xi^{supp}) \leq h(\xi^{supp})$ on an additional set of support points $\mathcal{X}_{support} = \{\xi_1^{supp}, \dots, \xi_N^{supp}\}$. We choose $\mathcal X_{support}$ to be a grid around the scenarios $\{\xi_1, ..., \xi_N\}$.
Ideally, the grid is very fine to enforce the requirement as much as possible. However, we observe that for our simple system the grid does not need to be very fine and increasing the number of grid points does not improve robustness anymore eventually. We use between $9$ and $49$ additional grid points in our reported experiments.

\subsection{Additional Experiments}

We report additional numerical experiments on the double integrator \eqref{eq:double_pend} and the exponential constraint in \eqref{eq:constraints}. Again, we make use of additional grid points, as outlined above. We visualize various distribution shifts $\lambda$ and regularizations $\epsilon$ for KDR-mpc in figure \ref{fig:kdrompc} and gradient regularization in figure \ref{fig:gradient}.

\begin{figure}[h!]
	\floatconts{fig:kdrompc}{\caption{KDR-mpc for a set of fixed epsilons. We create a model mismatch by adding a constant $\lambda$ to the dynamics matrix $A$. The red line is the constraint to be fulfilled.}}{
		\subfigure{
			\label{fig:kdro1}
			\includegraphics[width=0.3\textwidth]{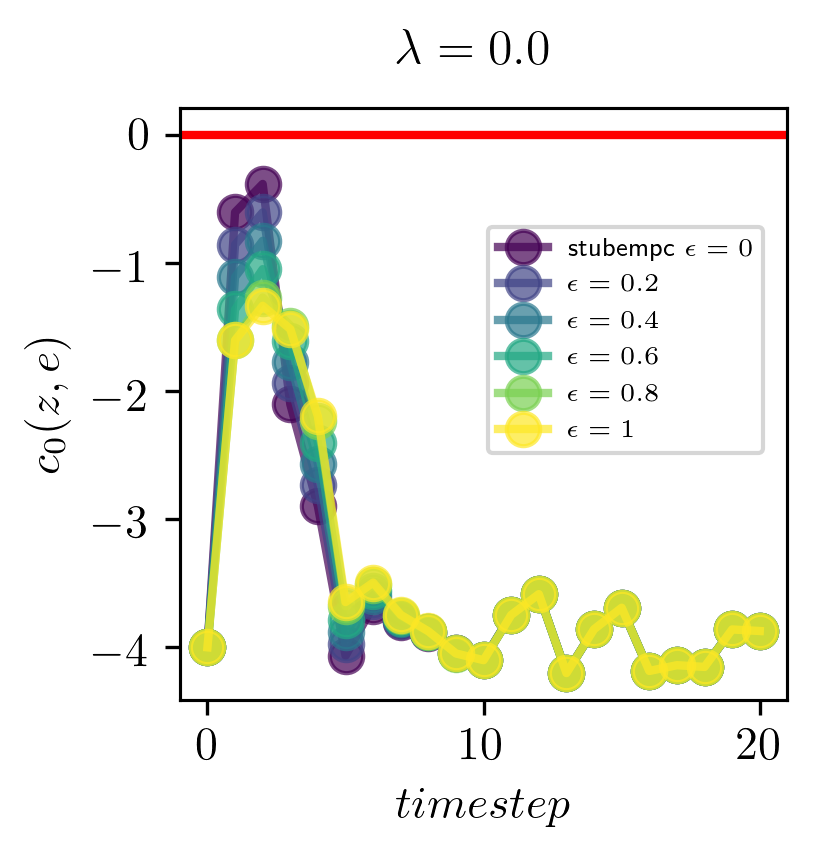}
		}
		\subfigure{
			\label{fig:kdro2}
			\includegraphics[width=0.3\textwidth]{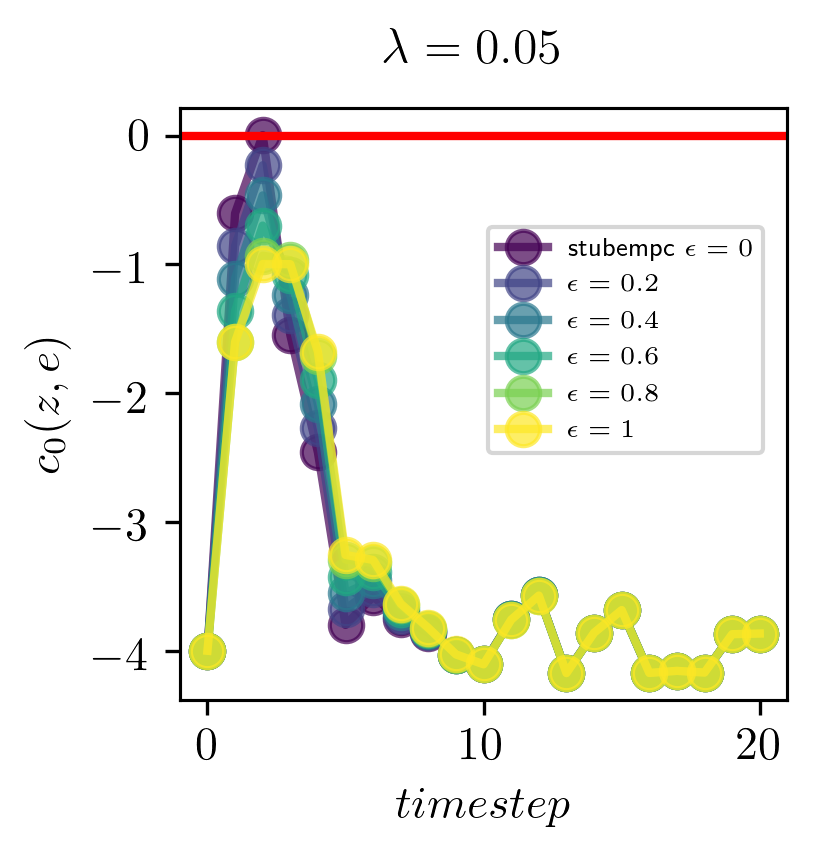}
		}
		\subfigure{
			\label{fig:kdro3}
			\includegraphics[width=0.3\textwidth]{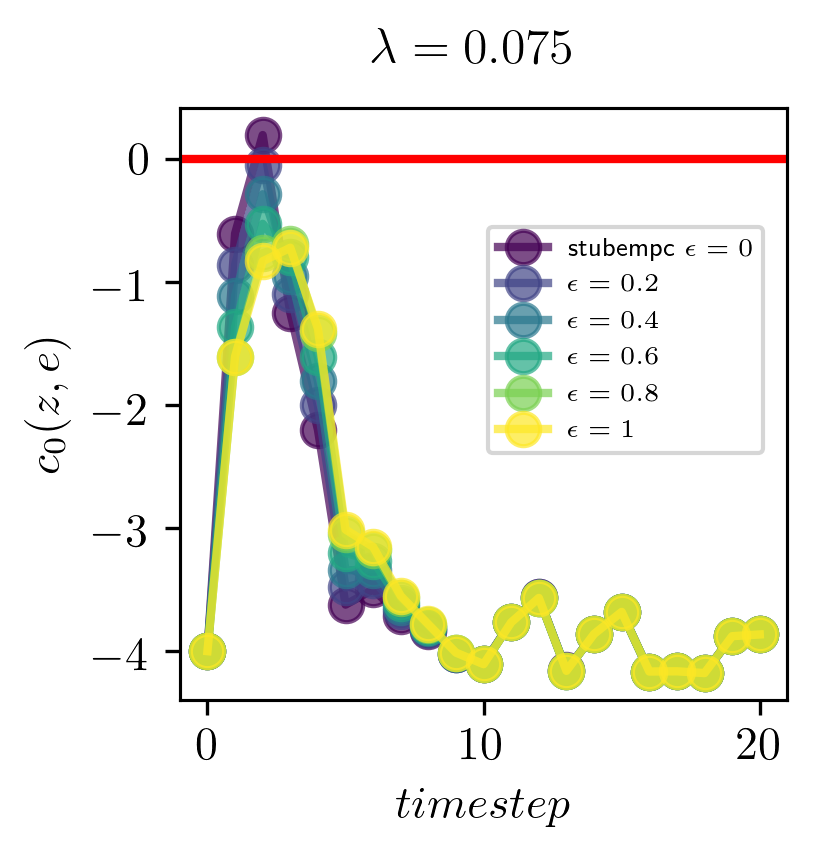}
		}\\
		\subfigure{
			\label{fig:kdro4}
			\includegraphics[width=0.3\textwidth]{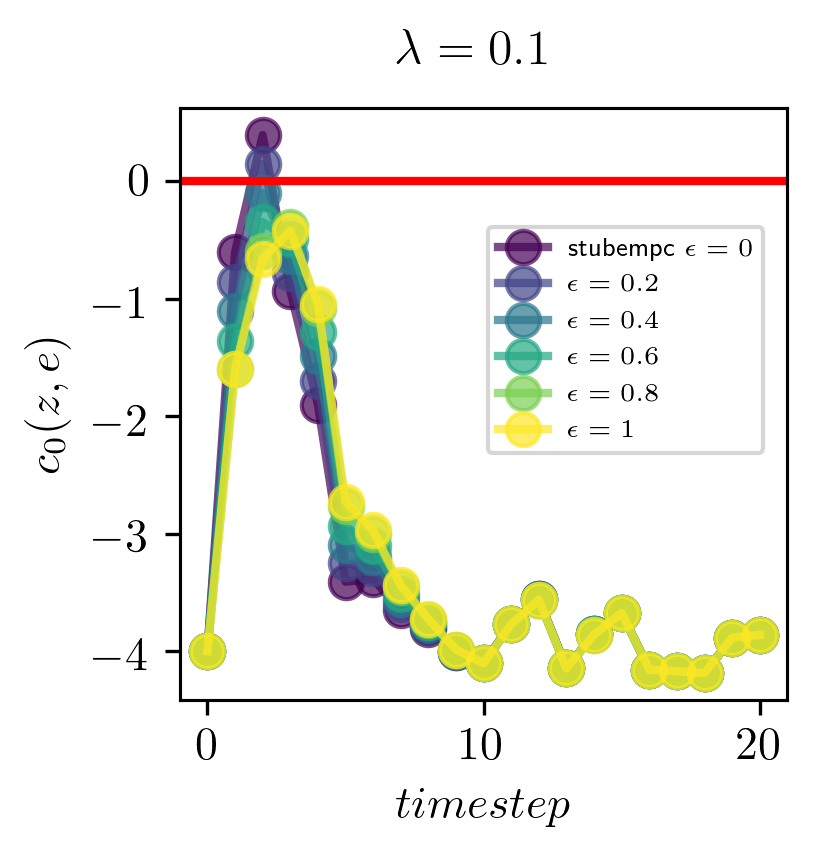}
		}
		\subfigure{
			\label{fig:kdro5}
			\includegraphics[width=0.3\textwidth]{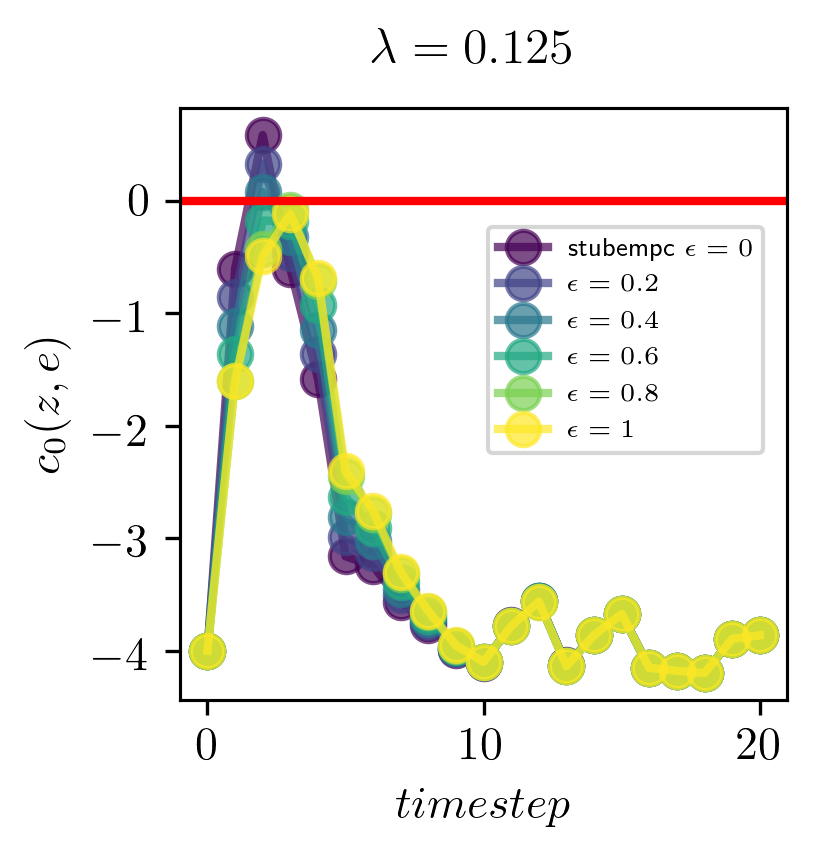}
		}
		\subfigure{
			\label{fig:kdro6}
			\includegraphics[width=0.3\textwidth]{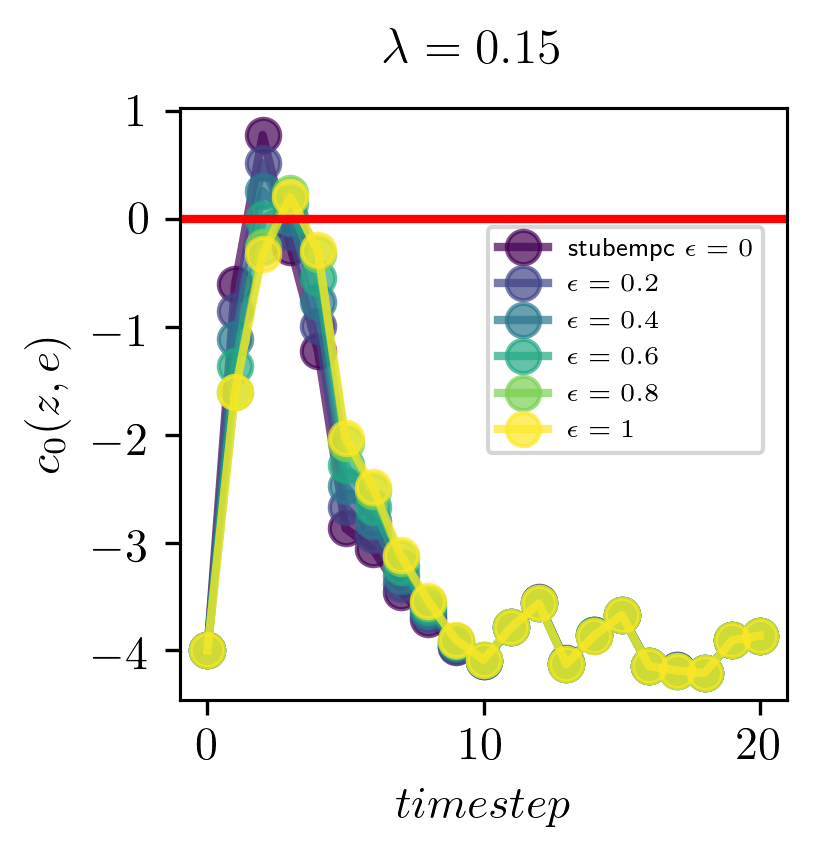}
		}
	}
\end{figure}

\begin{figure}[h!]
	\floatconts{fig:gradient}{\caption{
    Gradient regularization for a set of fixed epsilons. We create a model mismatch by adding a constant $\lambda$ to the dynamics matrix $A$. We can observe that eventually in fig. \ref{fig:pic6} the distribution shift gets too large to be able to satisfy constraints anymore even with additional robustness.
    }}{
		\subfigure{
			\label{fig:pic1}
			\includegraphics[width=0.3\textwidth]{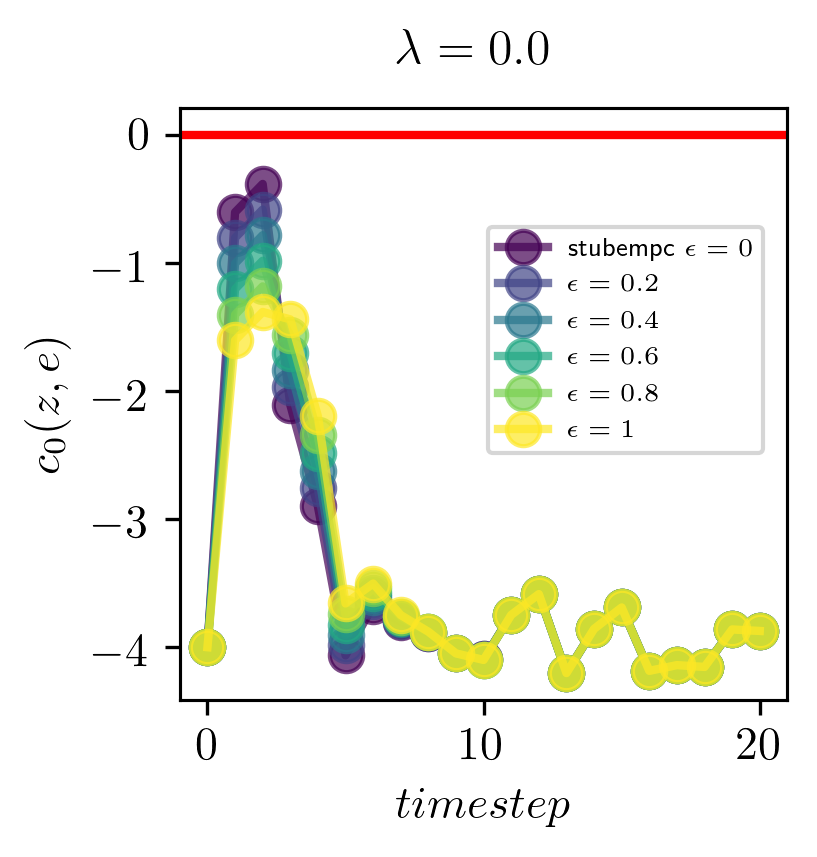}
		}
		\subfigure{
			\label{fig:pic2}
			\includegraphics[width=0.3\textwidth]{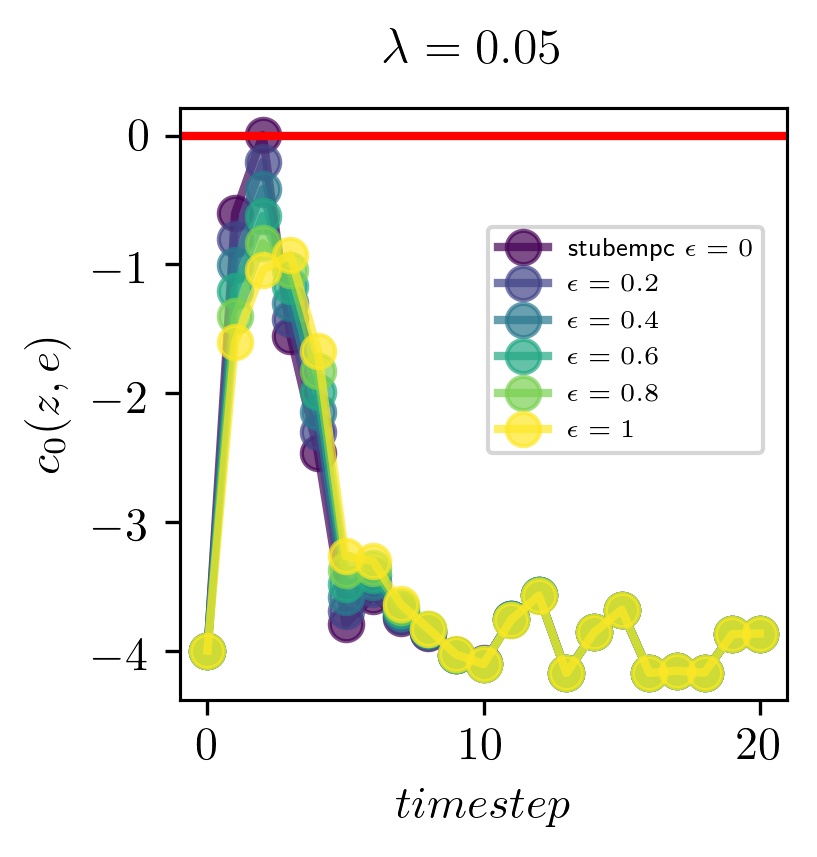}
		}
		\subfigure{
			\label{fig:pic3}
			\includegraphics[width=0.3\textwidth]{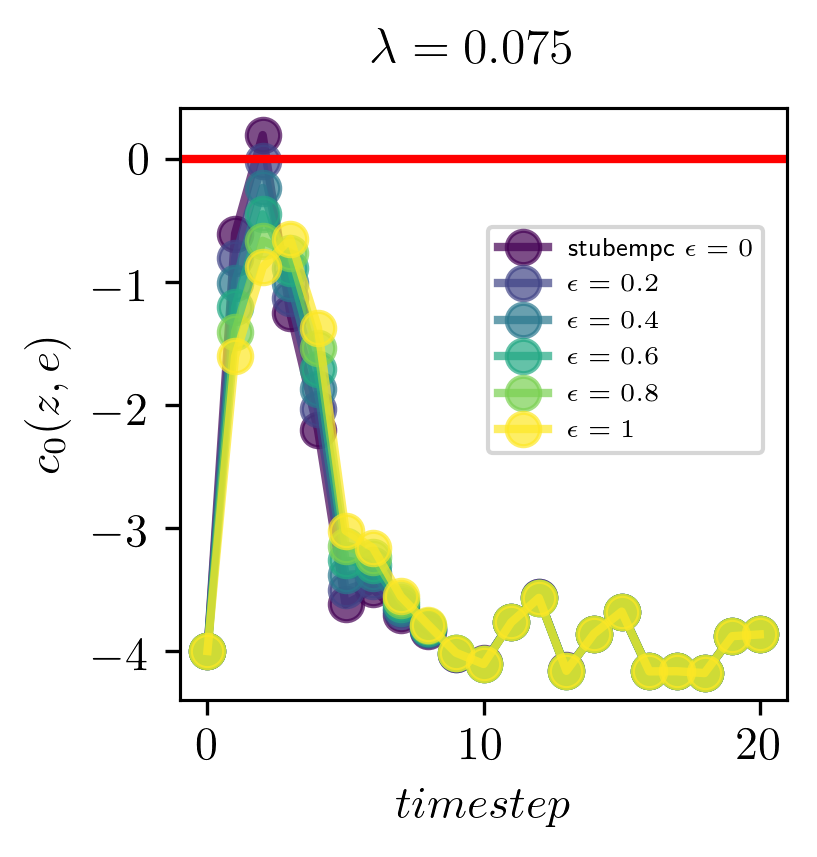}
		}\\
		\subfigure{
			\label{fig:pic4}
			\includegraphics[width=0.3\textwidth]{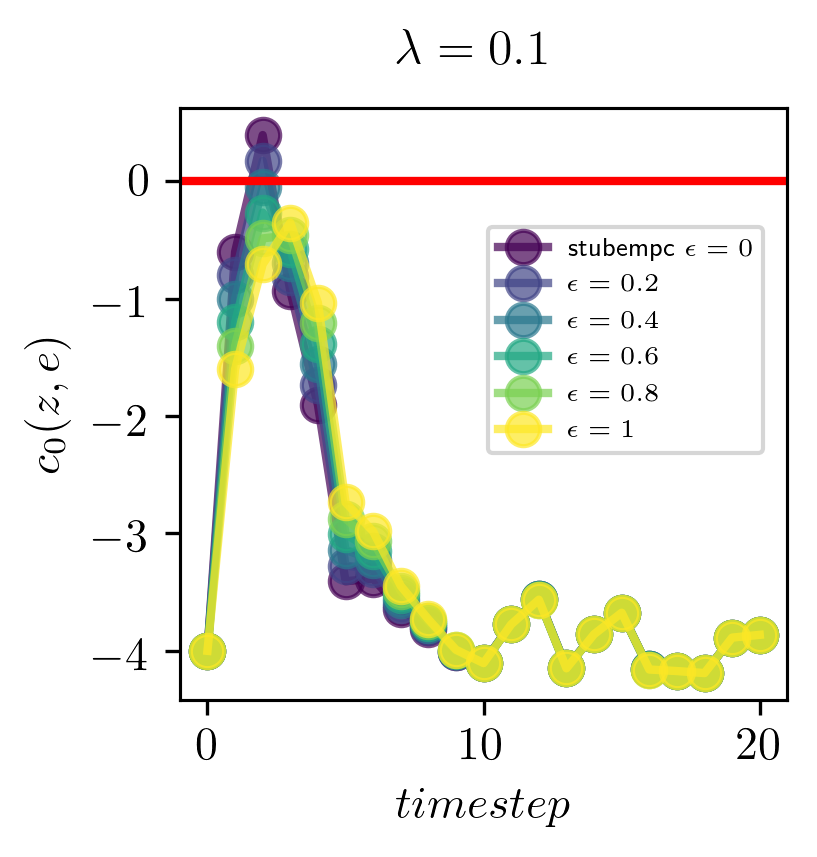}
		}
		\subfigure{
			\label{fig:pic5}
			\includegraphics[width=0.3\textwidth]{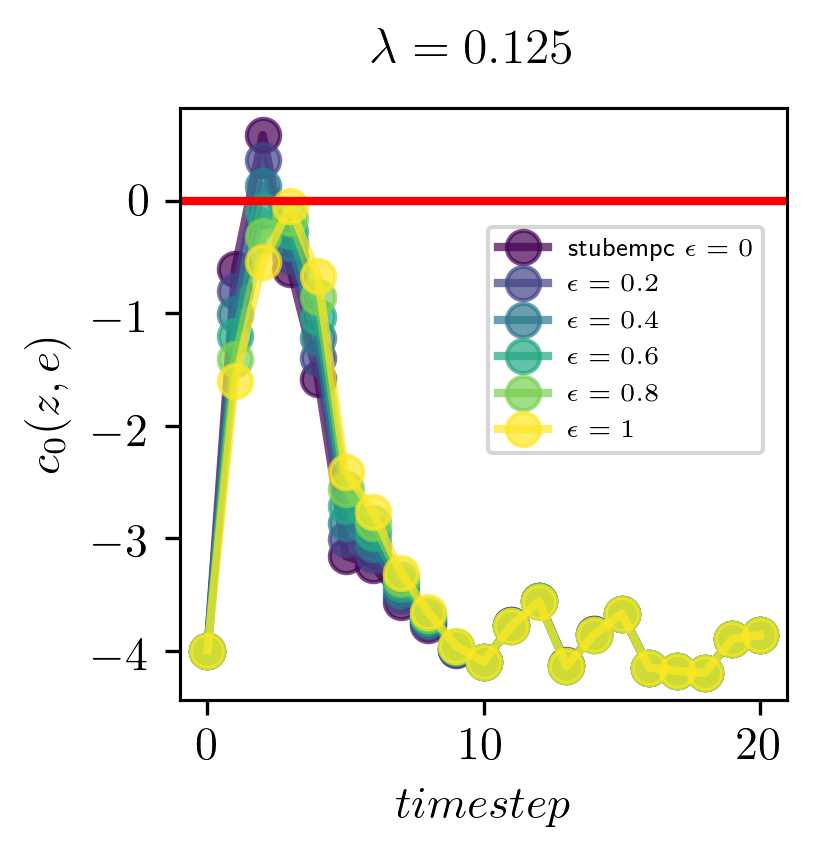}
		}
		\subfigure{
			\label{fig:pic6}
			\includegraphics[width=0.3\textwidth]{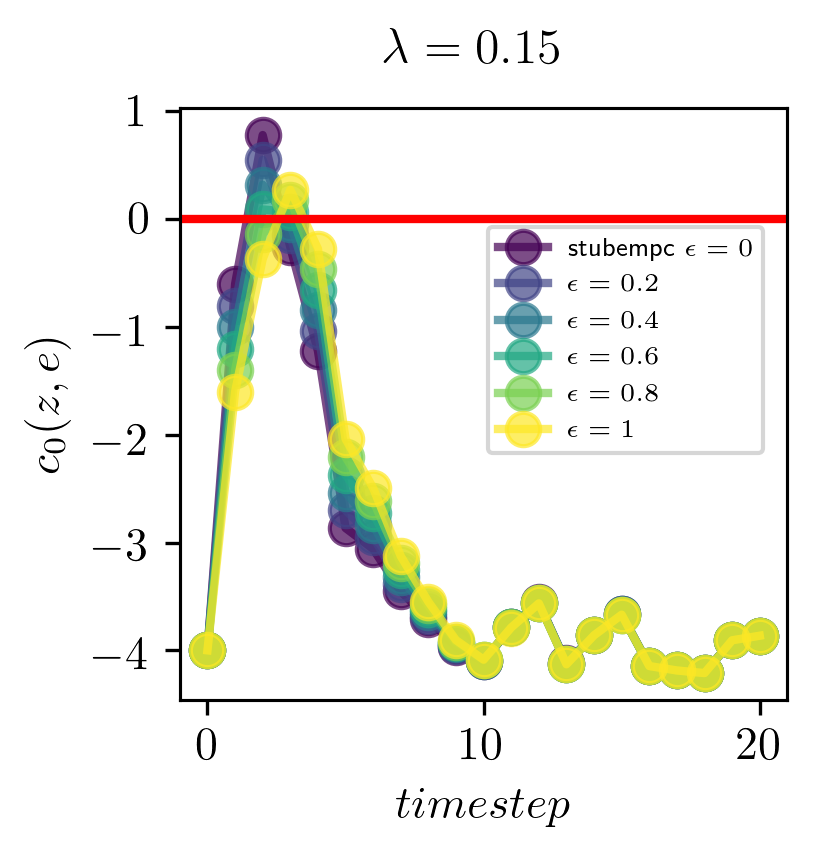}
		}
	}
  \end{figure}
\end{document}